\documentclass[9pt,twocolumn,twoside]{osajnl}

\journal{arxiv} 

\setboolean{shortarticle}{true}

\title{Engineering the spectral bandwidth of quantum cascade laser frequency combs}

\author[*,1]{Maximilian Beiser}
\author[*,1]{Nikola Opa{\v c}ak}
\author[1]{Johannes Hillbrand}
\author[1]{Gottfried Strasser}
\author[1,$\dagger$]{Benedikt Schwarz}

\affil[1]{Institute of Solid State Electronics, TU Wien, 1040 Vienna, Austria}

\affil[*]{These authors contributed equally to this work}
\affil[$\dagger$]{Corresponding author: benedikt.schwarz@tuwien.ac.at}

\begin{abstract}
Quantum cascade lasers (QCLs) facilitate compact optical frequency comb sources that operate in the mid-infrared and terahertz spectral regions, where many molecules have their fundamental absorption lines. Enhancing the optical bandwidth of these chip-sized lasers is of paramount importance to address their application in broadband high-precision spectroscopy. 
In this work, we provide a numerical and experimental investigation of the comb spectral width and show how it can be optimized to obtain its maximum value defined by the laser gain bandwidth.
The interplay of nonoptimal values of the resonant Kerr nonlinearity and the cavity dispersion can lead to significant narrowing of the comb spectrum and reveals the best approach for dispersion compensation. The implementation of high mirror losses is shown to be favourable and results in proliferation of the comb sidemodes. Ultimately, injection locking of QCLs by modulating the laser bias around the roundtrip frequency provides a stable external knob to control the FM comb state and recover the maximum spectral width of the unlocked laser state.
\end{abstract}

\setboolean{displaycopyright}{false}

\begin{document}

\maketitle

\section{Introduction}
The realization of optical frequency combs (OFCs)~\cite{haensch2006nobel,hall2006nobel} has lead to multiple breakthroughs in fundamental science and enabled manifold applications. The spectrum of frequency combs consists of a set of evenly spaced modes with a well-defined phase relation, which facilitates achievements like high precision spectroscopy down to attosecond resolution~\cite{Baltuka2003}, frequency synthesis and optical clocks~\cite{Udem2002Optical} and enhancing the search for exoplanets~\cite{Wilken2012}.
For spectroscopic applications, the mid-infrared spectral region is a molecular fingerprint region, since fundamental rotation-vibrational lines of many molecules are located there~\cite{keilmann2004time}. 
Conventional frequency comb sources in this region depend on a chain of optical components which limits their usability and potential for on-chip integration. To broaden the spectroscopic applications of such mid-infrared combs, a more compact and electrically driven frequency comb generator is preferred. 
Quantum cascade lasers (QCLs) are unipolar semiconductor lasers that emit in the mid-infrared and terahertz range, which became the dominant light sources in these spectral regions~\cite{faist1994quantum}.
In 2012, the first quantum cascade laser frequency comb was demonstrated by Hugi et al.~\cite{hugi2012mid}. Unlike the conventional OFCs that rely on the emission of short pulses, QCL combs produce frequency-modulated (FM) output with an almost constant intensity and a linear frequency chirp.  Moreover, the frequency comb regime in a  free-running QCL is self-starting and requires no additional optical elements, which led to broad attention from researches recently~\cite{Singleton2018,hugi2012mid}. 
Due to their on-chip integration, QCLs have vast potential to miniaturize high-precision spectroscopy.
Mid-infrared dual-comb spectroscopy based on QCLs~\cite{villares2014dual} was demonstrated and the viability of this technique for real-time monitoring of chemical reactions was proven~\cite{Klocke2020, Pinkowski2020}.
Furthermore, it was shown that injection of a microwave signal at the cavity roundtrip frequency enables locking of the repetition rate of the comb~\cite{StJean2014}, which allows coherent control of the FM comb states using established radio-frequency (RF) electronics~\cite{hillbrand2018coherent}. Recent research in this field demonstrated FM frequency comb formation in other laser types as well, e.g. interband cascade lasers~\cite{schwarz2019monolithic}, quantum dot lasers~\cite{Hillbrand2020inphase} and laser diodes~\cite{sterczewski2020}.
The missing essential step towards the broadband high-precision spectroscopic application of QCLs is a reliable control of the FM combs states and enhancement of their optical spectral bandwidth.

\begin{figure*}[t!]
	\centering
	\includegraphics[width =\textwidth]{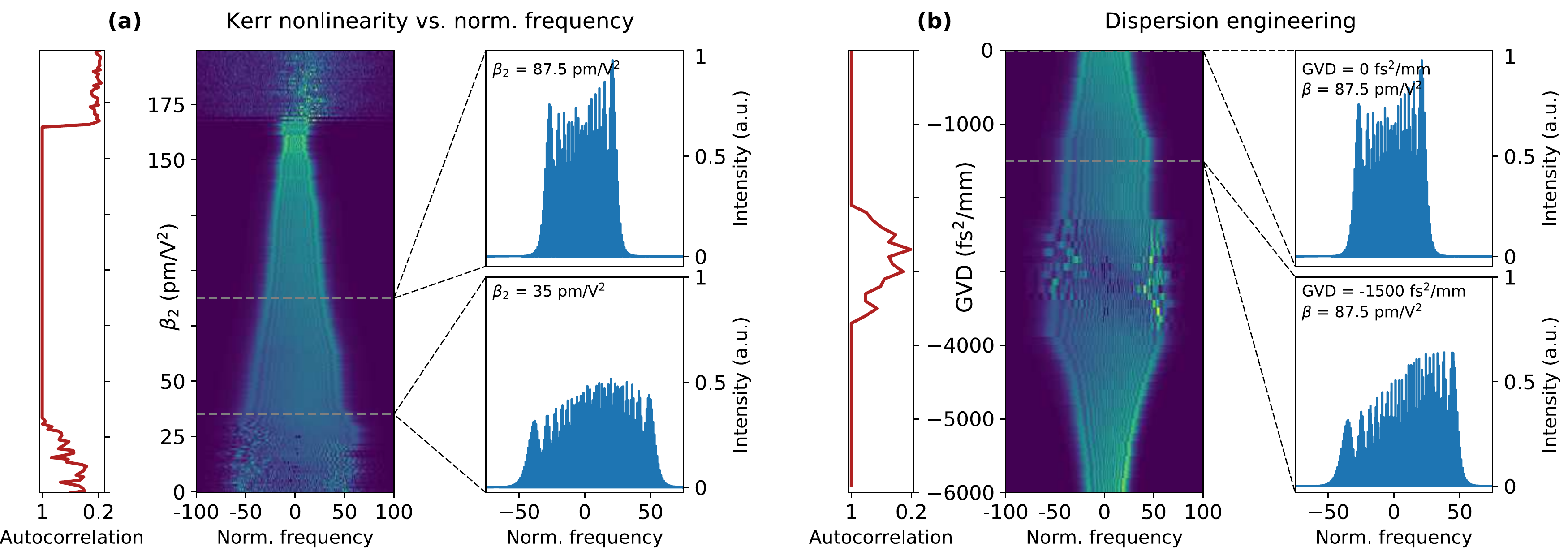}	
	\caption{ (a) Evolution of the intensity spectrum with the increase of the resonant Kerr nonlinearity $\beta_2$ for a dispersion compensated cavity (GVD = 0). Initial small values of $\beta_2$ lead to unlocked states, as is indicated by the autocorrelation value smaller than unity. FM comb formation appears between $\beta_2=$35 and 165 pm/V$^2$ with the autocorrelation equal to unity. Increase of the Kerr nonlinearity narrows the spectrum. Unlocked states are obtained above $\beta_2=$165 pm/V$^2$. Two insets on the right depict spectra taken at $\beta_2=$35 pm/V$^2$, with maximum spectral width and at 87.5 pm/V$^2$, with smaller number of comb modes. 
	(b) Evolution of the spectrum for $\beta_2=$87.5 pm/V$^2$ as the GVD is tuned. The maximum optical bandwidth is recovered for non-zero GVD of -1500 fs$ ^2$/mm.}
	\label{fig1_kerr}
\end{figure*}

The dispersion of QCL waveguides was early on identified to play an important role for FM comb formation.
A prevailing method to tune the group velocity dispersion (GVD) is the use of a Gires-Tournouis etalon, where a planar mirror is placed behind the facet of the laser to induce a wavelength dependent group delay~\cite{Hillbrand:18,Villares2016dispersion}. 
A different approach relies on the coupling of the dielectric waveguide to a plasmonic resonance in order to decrease the waveguide dispersion~\cite{Bidaux17}.
The experimental observations of FM combs in QCLs have mostly been found in lasers with rather small GVD, which led to the opinion that full dispersion compensation is the optimum.
However, recent theoretical work~\cite{art:Opacak19PRL,Burghoff:20} revealed the key role of the interplay of multiple effects in FM combs formation, such as the GVD, the Kerr nonlinearity, the intracavity light intensity profile, etc. 
It was shown, that the full compensation of the dispersion does not necessarily lead to broader comb spectra covering the entire available gain bandwidth or a comb operation at all.

\section{Discussion and Results}
In this paper, we show several ways of how FM combs can be influenced and optimized.
One of the main goals of FM comb engineering is the control of the number of comb modes in order to increase the optical bandwidth and ultimately facilitate a broad ruler for dual-comb spectroscopy.
The presented guideline explains the roles of the GVD and resonant Kerr nonlinearity with a numerical study and reveals the impact of the cavity facet reflectivities.
All sets of optimal parameters lead to the same maximum spectral width, which is predefined by the gainwidth of the laser active medium. We further present the theoretical and experimental proof that the same maximal width can also be achieved by RF modulation of the injected current. This constitutes an appealing alternative to spectral width optimization, which does not require a redesign or adaptation of the laser.
In a GVD compensated cavity the laser modes are innately equidistant, which seemingly facilitates mode-locking and frequency comb formation.
While this is the case for the traditional mode-locked lasers that emit short pulses, the situation is very different in case of FM comb operation. The linear frequency chirp observed for FM comb operation is the result of a complex synchronization phenomenon among the beatings of neighboring comb modes, analogous to anti-phase synchronization among coupled clocks~\cite{Hillbrand2020inphase}. Due to the ultra-fast gain recovery time of QCLs the spectral gain asymmetry due to Bloch gain yields a giant Kerr nonlinearity~\cite{art:Opacak19PRL,opacak2021bloch}. This induced resonant nonlinearity alters the optimal comb conditions, which is the reason why FM combs have mostly been found in GVD compensated cavities~\cite{Bidaux17}.
Figure~\ref{fig1_kerr} shows the dependence of the intensity spectrum on the Kerr non-linearity for a dispersion compensated QCL with uncoated identical facets. The simulation is based on the master equation, which considers spatial hole burning, dispersion and the Kerr nonlinearity~\cite{art:Opacak19PRL}. The laser was simulated to operate well above threshold at 60$\%$ of the maximum current.
Three different regimes can be distinguished depending on the Kerr non-linearity. Starting from zero nonlinearity, the laser is unlocked and emits a broad spectrum. Above a threshold minimum value of the nonlinearity (around $\beta_2$=35 pm/V$^2$), FM comb operation is obtained. Intriguingly, a further increase of the Kerr nonlinearity leads to a significant narrowing of the spectrum, while the laser remains in the locked state up to $\beta_2$=165 pm/V$^2$.
In the optimal case (35 pm/V$^2$), an FM comb is obtained with the broadest possible spectral width, while for a non optimal case (87.5 pm/V$^2$) the laser emits a much narrower spectrum.
As the Kerr nonlinearity originates from the resonant optical transitions and the asymmetry of the gainshape, it is determined by the band structure design~\cite{opacak2021bloch}. 
Consequently, further optimization would require a complete redesign of the active region in order to attain the optimal and maximal spectral width.
In the following we will discuss three alternative ways to recover the widest possible spectrum starting from the non-optimal active region (87.5 pm/V$^2$). 

Dispersion engineering can be applied to tune the laser towards the optimal FM operation. 
It was shown above, that a laser cavity with complete dispersion compensation is optimal for a nonlinearity of $\beta_2$=35 pm/V$^2$. However, for different nonlinearities a specific non-zero GVD value will result in the largest spectral span.
In figure ~\ref{fig1_kerr}b we display the numerical simulation results for the non-optimal Kerr non-linearity of $\beta_2$=87.5 pm/V$^2$, while sweeping the GVD from -6000$\,$fs$^2$/mm to zero. 
By tuning the GVD from zero to -1500$\,$fs$^2$/mm the same maximal spectral span can be recovered as in the case of $\beta_2$=35 pm/V$^2$. The interplay of non-zero GVD and the Kerr nonlinearity mimics the same effect on the FM comb spectral output.
A further increase of the GVD leads to an unlocked state and then again to a locked state with decreasing spectral span. 

\begin{figure}[t!]
	\centering
	\includegraphics[width = \linewidth]{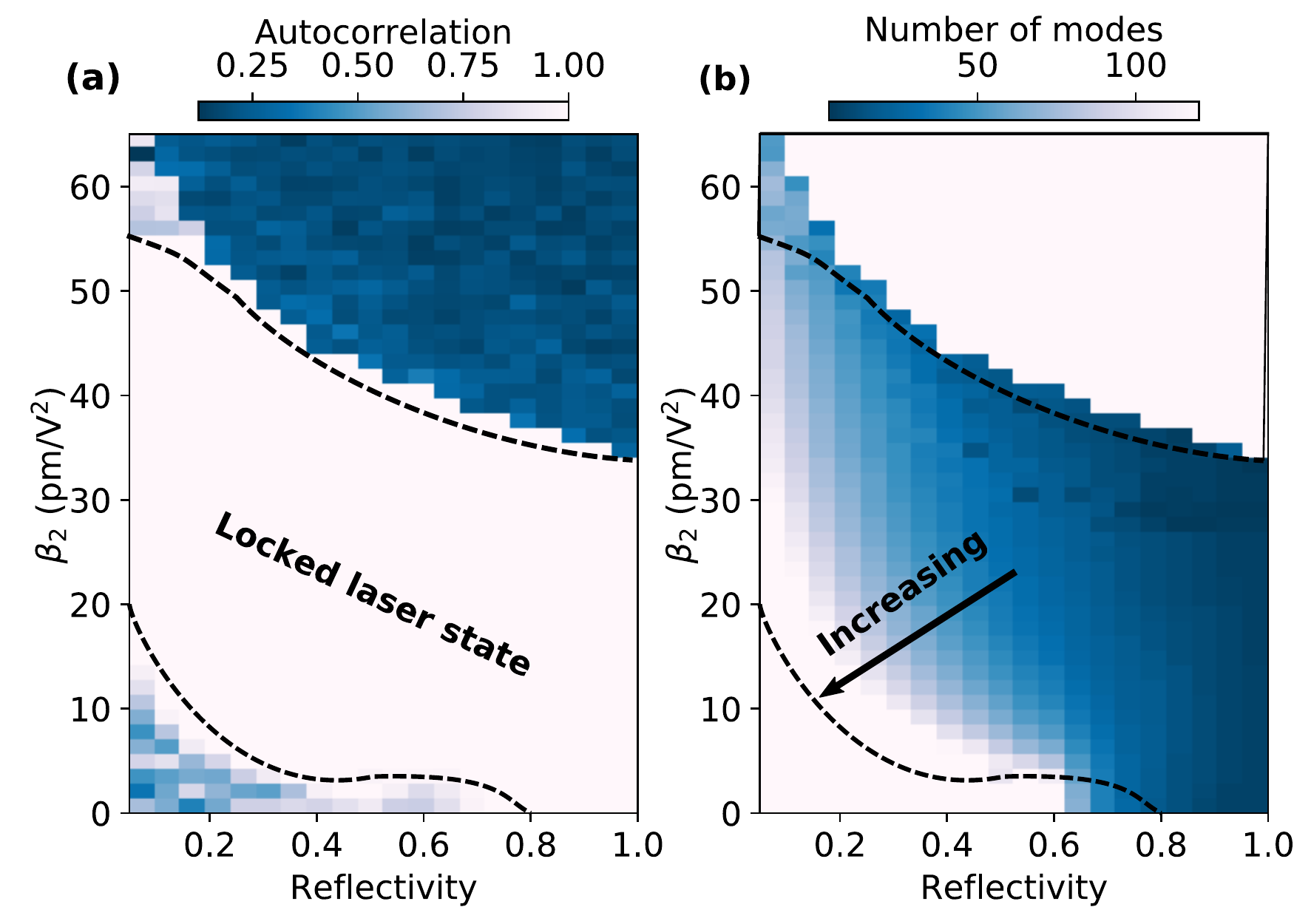}
	\caption{(a) Calculated autocorrelation as both the Kerr nonlinearity and the reflectivity of one cavity facet are tuned. Region of the parameter space where FM comb formation is obtained is marked with black dashed lines. (b) Calculated spectral width for the same parameter sweep. Optimal region with the largest number of comb modes is indicated with the arrow. Relevant data where an FM comb is obtained lies between the black dashed lines }
	\label{fig2_mergeplot}
\end{figure}

A novel second approach for comb engineering is the use of coatings on the facets, which changes their reflectivities.
In a Fabry-Perot laser cavity the intensities of the left and right propagating components of the field inside the cavity are increasing as they propagate, since the facets are semi-transparent. It has to be noted that the light intensity profile inside the cavity plays a crucial role in FM comb operation. The slope of the intensity has a large impact on the FM comb operation, which can consequently be altered by simply changing the coatings of the laser facets.
In figure \ref{fig2_mergeplot}a a colormap of the calculated autocorrelation of the laser field is displayed, where the value of one indicates a locked comb state. Conversely, unlocked states yield an autocorrelation factor smaller than unity. We sweep the reflectivity for one facet from zero to one with the other facet reflectivity set to one, while sweeping the Kerr nonlinearity from $\beta_2$=0 to 65 pm/V$^2$.

A large part of the parameter space is a locked FM comb state (indicated by the dashed lines). There are two areas of unlocked states when both the Kerr nonlinearity and the facet reflectivity are either low or relatively large. If the comb does not lock for low values of the nonlinearity, it can be brought into the locked regime by increasing the facet reflectivity. Conversely, for too high nonlinearities, the reflectivity needs to be decreased.
In figure \ref{fig2_mergeplot}b we show the corresponding spectral width in order to identify regions, where the comb is characterized with a broad intensity spectrum. Evidently, an optimal condition for a locked laser state with the broadest spectral output is obtained with a small reflectivity (close to the lower dashed line).
One can see, that the overlap for optimal conditions, a broad spectrum and locked laser state, is small compared to the size of the locking area in \ref{fig2_mergeplot}a. Furthermore, as the nonlinearity is determined by the design of the laser active region, only the reflectivity value can be tuned. Note that due to the interplay of the GVD and Kerr nonlinearity, dispersion engineering affects the comb state similarly as changing the nonlinearity in \ref{fig2_mergeplot}a. 
Our results emphasize that changing the reflectivity of the laser facet can serve as a practical way to impact the FM comb operation, particularly as coating technology is already a part of laser fabrication.

\begin{figure*}[t!]
	\centering
	\includegraphics[width =\textwidth]{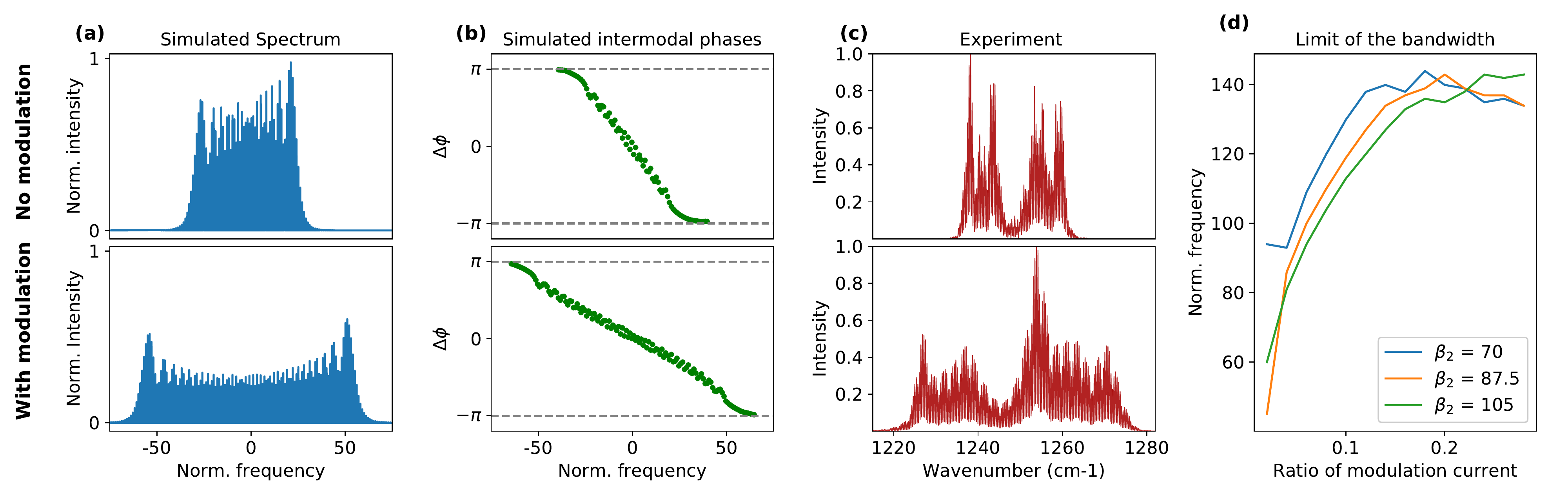}
	
	\caption{Simulated (a) intensity spectrum and (b) intermodal phase differences with (bottom) and without (top) the modulation of the injection current. (c) Experimental intensity spectrum with (bottom) and without (top) the modulation of the injection current.  (d) Simulated spectral width depending on the modulation depth.  }
	\label{fig3_modulation}
\end{figure*}
As a third approach we present a technique which does not require a precise adjustment of the cavity for a specific active region -- the injection of an RF signal around the cavity roundtrip frequency.
In order to model the influence of the modulation of the injection current on the spectrum of FM combs, we again employ the coherent master equation~\cite{art:Opacak19PRL, Hillbrand2020shortpulses}. We furthermore provide an experimental proof, similar to recent work~\cite{kapsalidis2020midinfrared}. The laser has the same non-optimal Kerr nonlinearity of 87.5 pm/V$^2$ and zero GVD. The current modulation is applied to a short section which covers 10\% of the laser cavity.
In figure \ref{fig3_modulation}a\&b we show the intensity spectrum and the phase differences of neighboring modes of the FM comb, both analyzed with and without the current modulation.
Without modulation, the comb spectrum  spans over 74 modes. The intermodal phase differences are linearly splayed over the full range of $2\pi$, which is characteristic for FM combs and results in a linear frequency chirp. The lower plot of figure \ref{fig3_modulation}a shows the optical spectrum, when the laser is subjected to strong current modulation with an amplitude of 10$\%$ of the DC driving current. With RF modulation the comb spectrum significantly broadens from 74 to 134 modes. The slope of the intermodal phase differences in figure \ref{fig3_modulation}b broadens accordingly and flattens out, demonstrating the locked coherent behaviour of the FM comb state. 
We experimentally investigated the predicted behaviour with a high performance mid-infrared quantum cascade laser (QCL) frequency comb emitting at 8$\mu$m. The QCL cavity is split into a 3.15 mm long gain section and a 350 $\mu$m short modulation section optimized for efficient RF injection. Besides the optimized RF geometry of the modulation section, the band structure of the QCL active material was designed as a single stack bifunctional active region~\cite{Schwarz2012bifunctional, Hillbrand2020shortpulses}. When operated at zero-bias, the QCL works as detector at its lasing wavelength and thus induces strong optical absorption. This allows a more efficient modulation of the laser with larger modulation depths.
In order to study the response of the QCL combs spectrum, we operate the modulation section at a DC bias of 2.8V to maximize the modulation efficiency.
The modulation depth is increased continuously up to 35 dBm injection power. The associated optical output spectrum is displayed in figure \ref{fig3_modulation}c. 
The spectrum of the free running QCL shows an optical bandwidth of 30 cm$^{-1}$ without any modulation. When injecting RF power at the cold cavity roundtrip frequency, the spectrum is broadened to 50 cm$^{-1}$. This experimental proof shows that RF injection can not only be used to stabilize the FM comb~\cite{hillbrand2018coherent}, but also to optimize the FM comb state and recover its maximum spectral width, which is defined by the laser gainwidth.
All three presented methods gave a similar spectral width under optimal conditions for the FM comb.
To further explore this upper limit, we provide a more detailed study of the spectral broadening by RF modulation.
We simulated the spectral broadening depending on the modulation depth ($I_{RF}/I_{DC}=0 - 0.3$) for fixed values of the Kerr non-linearity of 70, 87.5 and 105 pm/$V^2$ and an injection frequency 10 MHz below the the cold cavity roundtrip time. The results are displayed in figure \ref{fig3_modulation}d. For all values of the nonlinearity, the simulated behaviour is strikingly similar.
The spectral width increases considerably up to a modulation depth of 0.15, where it saturates at an approximately constant bandwidth of 140 modes.
The maximal optical bandwidth achievable by RF modulation is practically equal to the width of the unlocked state. This result shows that the spectral width of the FM comb does not exceed the width of the unlocked laser~\cite{kapsalidis2020midinfrared,Khurgin2020analytical}. 

\section{Conclusion}
In conclusion, we provided a guideline to engineer FM combs for broadband spectroscopic application. We investigated the influence of multiple cavity parameters of FM combs by using numerical models as well as experiments. 
Due to the inherent resonant Kerr nonlinearity of the QCL active region, strictly reducing and compensating the GVD does not result in an optimal frequency comb regime, in contrast to previous expectations.
The profile of the intracavity light intensity has a crucial role in the FM comb formation. It can be controlled by tuning the laser facet reflectivities in order to broaden the comb spectra. Moreover, we demonstrate that high mirror losses are preferable. 
Employing a sophisticated model using the master equation approach, we simulate the FM comb response to a strong modulation of the injection current around the cavity roundtrip frequency. The experiment with an injection locked high performance bifunctional quantum cascade laser frequency comb displays excellent agreement with the modelled behaviour. 
To overcome the restrictions that are dominantly imposed by the laser design, RF injection is the method of choice to circumvent the change of the laser active region and achieve a broad spectral output with reasonable effort. 
The further use of RF modulation as an efficient technique to broaden the spectrum of FM comb will enhance the QCL applications in dual-comb spectroscopy.

\begin{backmatter}
\bmsection{Funding}
This project has received funding from the European ResearchCouncil (ERC) under the European Union’s Horizon 2020 researchand innovation programme (Grant agreement No. 853014). 
\bmsection{Disclosures}
The authors declare no conflicts of interest.

\end{backmatter}

\bibliography{sample}

\bibliographyfullrefs{sample}

\end{document}